\documentclass{easychair}

\usepackage{doc}
\usepackage{makeidx}
\usepackage{amsmath, amsthm, amssymb}
\usepackage{array,color,colortbl}

\usepackage{epsf}
\usepackage{epstopdf}
\include{psfig}
\input epsf

\input epsf

\begin{document}

%
\title{Populations of models, Experimental Designs and coverage of parameter space by Latin Hypercube and Orthogonal Sampling}


\titlerunning{A comparison of coverage of parameter space for populations of models}

%
\author{
Kevin Burrage\inst{1}
\and
    Pamela  Burrage\inst{2}
\and
   Diane Donovan\inst{3}
\and\\
   Bevan Thompson\inst{3}
}

\institute{
  Department of Computer Science, University of Oxford, UK; Mathematical Sciences,
 Queensland University of Technology, Queensland 4072,
Australia. \\ \texttt{kevin.burrage@gmail.com}\\
\and
  Mathematical Sciences,
 Queensland University of Technology, Queensland 4072,
Australia. \\
\texttt{pamela.burrage@qut.edu.au}\\
\and
   School of Mathematics and Physics, The University of Queensland,
 Queensland 4072,
Australia. \\
\texttt{dmd@maths.uq.edu.au, hbt@maths.uq.edu.au}\\
 }

\authorrunning{Burrage, Burrage, Donovan and Thompson}

\clearpage

\maketitle

\begin{abstract}
In this paper we have used simulations to make a conjecture about the coverage of a $t$ dimensional subspace of a $d$ dimensional parameter space of size $n$ when performing $k$ trials of Latin Hypercube sampling. This takes the form $P(k,n,d,t)=1-e^{-k/n^{t-1}}$. We suggest that this coverage formula is independent of $d$ and this allows us to make connections between building Populations of Models and Experimental Designs.  We also show that Orthogonal sampling is superior to Latin Hypercube sampling in terms of allowing a more uniform coverage  of the $t$ dimensional subspace at the sub-block size level.
\end{abstract}

\setcounter{tocdepth}{2}

%
%

\pagestyle{empty}

\section{Introduction}
\label{sect:introduction}

Mathematical models are frequently highly tuned with parameters being given to many decimal places. These parameters are often fitted to a set of mean observational/experimental
data and so the inherent variability in the underlying dynamical processes is not captured.
A very recent approach for capturing this important and intrinsic variability is based around the concept of a population of models (POMs) \cite{MT} in which a mathematical model is built that has a set of points,  rather than a single point, in parameter space,  all of which are selected to fit a set of experimental/observational data.

Since first proposing the POM approach for neuroscience modelling, it has been extended to cardiac electrophysiology  \cite{A1}, \cite{A2}.  In that setting, biomarkers, such as Action Potential Duration and beat-to-beat variability, are extracted from time course profiles and then the models are calibrated against these biomarkers. Upper and lower values of each biomarker as observed in the experimental data are used to guarantee that estimates of variability are within biological ranges for any model to be included in the population. If the data cannot be characterised by a set of biomarkers then time course profiles can be used and a normalised root-mean-square (NRMS) comparison between the data values and the simulation values at a set of time points can be used to calibrate the population.

This approach suggests a possible new way in which science is done.  Firstly, the POM approach leads to methodologies that are essentially probabilistic in nature.  Secondly, it gives greater weight to the experimentation, modelling, simulation feedback paradigm \cite{A3}.
By implementing experiments based on a population of models, as distinct from experiments based on a single model, the  variability in the underlying structure can be captured by allowing changes in the parameters values. Such an approach avoids complications arising from  decisions on the use of ``best'' or ``mean'' data, and  the difficulties of identifying such data.

Building populations of models requires the generation of a number of parameter sets for the initial population, sampled from a possibly high-dimensional parameter space. With recent advances in computational power, it is possible to generate large numbers of such models,  leading to a better understanding  of the  systems under investigation. There are many ways to sample the parameter space, depending on costing constraints and therefore limits of computation.  A parameter sweep will cover the whole parameter space at a certain discrete resolution, while random sampling, Latin Hypercube sampling (LHS) and Orthogonal sampling (OS) will give increasingly improved coverage of parameter space when the number of samples is fixed and independent of the dimension of the space.

 In this paper we focus on LHS, a technique first introduced by  McKay, Beckman and Conover  \cite{MBC}. Suppose that the $d$ dimensional parameter space is divided into $n$ equally sized subdivisions in each dimension. A Latin Hypercube (LH) trial is a set of $n$ random samples with one  from each  subdivision; that is, each sample is the only one in each axis-aligned hyperplane containing it. McKay, Beckman and Conover suggested that the advantage of LHS is that the values for each dimension are fully stratified, while Stein \cite{stein} showed that  with LHS there is a form of variance reduction compared with uniform random sampling.  A variant of LHS, known as orthogonal sampling, adds the requirement that the entire sample space must be sampled evenly.

Depending on the underlying application POMs may be constructed in a number of different ways.  In \cite{A1}, \cite{A2}, for example, POMs are developed from  LHS, a useful approach because it provides insights into the nature of variability in cardiac electrophysiology.  In this case the coverage of parameter space, as long as it is random in some appropriate manner, is less of an issue than for the case where POMs are  used for parameter fitting. In this setting, POMs have similarities with Approximate Bayesian Computation (ABC) \cite{A5}.  By contrast, in ABC the sampling is usually performed adaptively so as to converge to subregions of parameter space where the calibrated models lie, as distinct from random sampling of the entire space. Thus in certain circumstances,  it is important to estimate the expected coverage of parameter space given $k$  Latin Hypercube trials of $d$-dimension.

In the paper \cite{A4} the authors  focused on  estimating the expected coverage of a $2$-dimensional parameter space for a population of $k$ LH trials with each trial of size $n$. In particular, counting arguments were used to predict  the expected coverage  of points in the parameter space after $k$ trials. These estimates were compared against numerical results based on a MATLAB implementation of  100 simulations. The results of the simulations led the authors to conjecture that the expected percentage coverage by $k$ trials of a $2$-dimensional parameter space, over values $1,2,\dots,n$,   tended to $1-e^{-k/n}$.

As McKay, Beckman and Conover \cite{MBC} state an advantage of LHS is that it stratifies each univariate mean simultaneously. Tang \cite{A7} and others have suggested that it may also be important to stratify the bivariate margins.  For instance, an experimental design  may involve a large number of variables, but in reality only a relatively small number of these variables are virtually effective. One way of dealing with this problem has been to project the factors onto a subspace spanned by the effective variables. However this can result in a replication of sample points within the effective subspace. Welch et al. \cite{A8}  suggest  LHS as a method for screening for effective factors, but Tang notes that there is still no guarantee that even in the  case of bivariate margins that this projection is uniformly distributed. Thus as an alternative, Tang \cite{A7} advocates Orthogonal sampling  and proposes a technique based on the existence of orthogonal arrays. He goes on to show that  Orthogonal sampling achieves uniformity on small dimensional margins and further that there is a form of variance reduction. Tang's approach is to start with an orthogonal array (defined in Section 2) and to replace its entries by random permutations to obtain an orthogonal sample. We will expand on this idea in Section 2, as well as describing an alternate method for Orthogonal sampling.

Orthogonal arrays and covering arrays have  been used also for generating interaction test suites for the testing of component-based systems. It is recognised that for large systems exhaustive testing may not be feasible, and instead suites are designed to test for  $t$-way interactions, for $t=2,\dots, 6$;  for details see \cite{BC2}, \cite{KR}. In \cite{BC1} and \cite{BC2}, Bryce and Colbourn give a density based greedy algorithm for the generation of  covering arrays for testing $t\geq 2$ interactions. This research and that in \cite{A3} have led us to investigate the  relationship between Experimental Design and  building POMs.

Thus in this paper rather than focusing solely on the coverage of the $d$-dimensional parameter space we wish to investigate the coverage of these lower dimensional subspaces. The motivation for this  is that resource constraints restricting the size of the population of models may preclude significant coverage of the entire parameter space. However, it may be desirable to know if such a population of models calibrates  for interactions of ``small strength''  by checking for all possible combinations of levels for, say, pairs or triples of variables. This would equate to investigating  the coverage of two and three dimensional subspaces. The justification is that statistical techniques may be  used to compare results   for pairwise or three-way interactions.  We will approach this question through the use of both LHS and OS.

In Section 2, we will give further discussion on LHS and OS.  We will also briefly review  Tang's construction for Orthogonal sampling and then given an alternate method for the generation of Orthogonal samples. In Section 3 we report on  MATLAB implementations of simulations of Latin Hypercube trials and Orthogonal sampling to test for uniform coverage of lower dimensional subspaces. In Section 4 we discuss and summarise the results from Section 3  as well as discussing future directions.

 \section{The construction of orthogonal samples}

Before introducing constructions we review the well known methods used to generate  Latin Hypercube samples and formalise the definitions for Orthogonal Samples.

 A Latin Hypercube trial generates an $n$ by $d$ matrix where each column is a random permutation of $\{1,2,\dots,n\}$ and then each row forms a $d$-tuple of the Latin Hypercube trial.  Thus given an experiment on $d$ variables each taking parameter values  $1,2,\dots,n$, a {\em Latin Hypercube trial}  is a randomly generated subset of $n$ points from a $d$-dimensional parameter space  satisfying the condition that the projections onto each of the $1$-dimensional subspaces are permutations; so for each variable the $n$ points  cover all possible parameter values for the corresponding subspace. By way of an example we take $d=3$ and $n=8$ giving below two Latin Hypercube trials LHS1 and  LHS2.

 \begin{align*}
 \begin{array}{c|c|c|c}
 \mbox{ LHS}1&\mbox{ LHS}2&\mbox{ LHS}3& \mbox{ OS LHS}4\\
\begin{array}{ccc}
(1,&2,&1)\\
(2,&3,&3)\\
(3,&1,&2)\\
(4,&7,&8)\\
(5,&8,&5)\\
(6,&5,&4)\\
(7,&4,&6)\\
(8,&6,&7)\\
\end{array}&
\begin{array}{ccc}
(1,&3,&2)\\
(2,&4,&6)\\
(3,&5,&3)\\
(4,&7,&8)\\
(5,&1,&1)\\
(6,&2,&7)\\
(7,&8,&4)\\
(8,&6,&5)\\
\end{array}&
\begin{array}{ccc}
((1,1),&(1,2),&(1,1))\\
((1,2),&(1,3),&(1,3))\\
((1,3),&(1,1),&(1,2))\\
((1,4),&(2,3),&(2,4))\\
((2,1),&(2,4),&(2,1))\\
((2,2),&(2,1),&(1,4))\\
((2,3),&(1,4),&(2,2))\\
((2,4),&(2,2),&(2,3))\\
\end{array}&
\begin{array}{ccc}
((1,1),&(1,3),&(1,2))\\
((1,2),&(1,4),&(2,2))\\
((1,3),&(2,1),&(1,3))\\
((1,4),&(2,3),&(2,4))\\
((2,1),&(1,1),&(1,1))\\
((2,2),&(1,2),&(2,3))\\
((2,3),&(2,4),&(1,4))\\
((2,4),&(2,2),&(2,1))\\
\end{array}
\end{array}
\end{align*}

Formally, a Latin Hypercube trial  $H$ is said to be an {\em Orthogonal Sample} (OS) if $n=p^d$  and for each of the $p^d$ $d$-tuple of the form $(p_1,p_2,\dots,p_d)$, where $1\leq p_i\leq p$, there exists an element of $H$ of the form
 $((p_1,x_1),(p_2,x_2),\dots,(p_d,x_d))$, where $1\leq x_i\leq p^{d-1}$. In the above examples, $d=3$ and $n=8=2^3$ thus $p=2$ and $p^{d-1}=4$. So we rewrite the numbers $1,\dots, n=8$ as $1\sim (1,1)$, $2\sim (1,2)$, $3\sim (1,3)$, $4\sim (1,4)$, $5\sim (2,1)$, $6\sim (2,2)$, $7\sim (2,3)$  and $8\sim (2,4)$. Using this representation we rewrite LHS1 as LHS3 and LHS2 as LHS4. Consider LHS3 and take the first two $3$-tuples $((1,1),(1,2),(1,1))$ and $((1,2),(1,3),(1,3))$ and project each ordered pair onto its first coordinate; that is, $((1,1),(1,2),(1,1))\longrightarrow (1,1,1)$ and $((1,2),(1,3),(1,3))\longrightarrow (1,1,1)$. Then in  the eight $3$-tuples of LHS3 we see $(1,1,1)$ twice, and so we can not get all distinct eight $3$-tuples on the set $\{1,2\}$. Therefore LHS3 is not an orthogonal sample, however we can check that LHS4 is an Orthogonal Sample.

Tang's \cite{A7} construction for Orthogonal Samples is based on the existence of orthogonal arrays.  These are structures that can be generalised to covering arrays. An {\em orthogonal array} OA$(N,d,n,t)$ on $d$ factors, of strength $t,$  over the set $X=\{1,2,\dots,n\}$ is a subset of the $d$-dimensional space $\overbrace{X\times X\times \dots \times X}^{d\ times}$ with the property that the projection onto any of the ${d\choose t}$ $t$-dimensional subspace $\overbrace{X\times X\times \dots \times X}^{t\ times}$ covers the entire subspace with multiplicity $\lambda$, where $N=\lambda n^t$. In a  {\em covering array}  the projections onto all $t$-dimensional subspaces   cover the entire subspace with multiplicity at least  $\lambda$.

  Tang takes a random orthogonal array and replaces each value $x$, $1\leq x\leq n$, by an $n\times 1$ vector where the entries correspond to a random permutation on the set $\{(x-1)\lambda n^{t-1}+1, \dots, (x-1)\lambda n^{t-1}+\lambda n^{t-1}=x\lambda n^{t-1}\} $. The rows of this new $n^{t+1}\lambda\times d$ form the tuples of a Latin Hypercube trial which is also an orthogonal sample. The random orthogonal array is achieved by taking an orthogonal array and randomly permuting rows, columns and values within a column.

  By contrast we have constructed $d$-dimensional orthogonal samples (where variables take the values $1,\dots, n=p^d$, for some positive integer $p$) using the following procedure:
\vspace{5mm}

\noindent {PROCEDURE:}
  \begin{itemize}
  \item Open  an  $p^d\times 2d$ array $A=[a(i,j)]$ and an $p^d\times d$ array $B=[b(i,j)]$.
  \item Generate all possible $p^d$ $d$-tuples with entries chosen from $1,\dots, p$.
  \item Assign each $d$-tuple to a separate row of $A$. Then if
     $(p_1,p_2,\dots, p_d)$ is assigned to row $i$,  set $a(i,2j-1)=p_j$,  $1\leq j\leq d$.
      \item For each $1\leq j\leq d$ columns $2j-1$ and $2j$ are filled as follows. For each $1\leq x\leq p$, identify all rows $i$ such that  $a(i,2j-1)=x$. Note that there are $ p^{d-1}$ rows for each $x$. Generate a random permutation on the set
      $\{1,\dots, p^{d-1}\}$ and assign these values sequentially to the $p^{d-1}$ entries $a(i,2j)$.
      \item For $1\leq i\leq p^d$ and $1\leq j\leq d$ set $b(i,j)=(a(i,2j-1)-1)p+a(i,2j)$.
 \end{itemize}
It is now easy to check that $B$ satisfies the definition of a Latin Hypercube trial and also an Orthogonal sample.

 \section{Simulations}

 In \cite{A4} we used Matlab simulations to make conjectures about the coverage of parameter space in terms of the number $k$ of Latin Hypercube trials  given the variable size $n$ for dimension $d=2$. In this section we look at the coverage of $t=2$ and $t=3$ dimensional subspaces in the $d=3,4,5$ dimensional parameter space.

\begin{figure}[htb]
\caption{Coverage for 2-tuples (left) and 3-tuples (right), for LHS, $d = 3$}
\centering
$$\begin{array}{cc}
\includegraphics[width=0.45\textwidth]{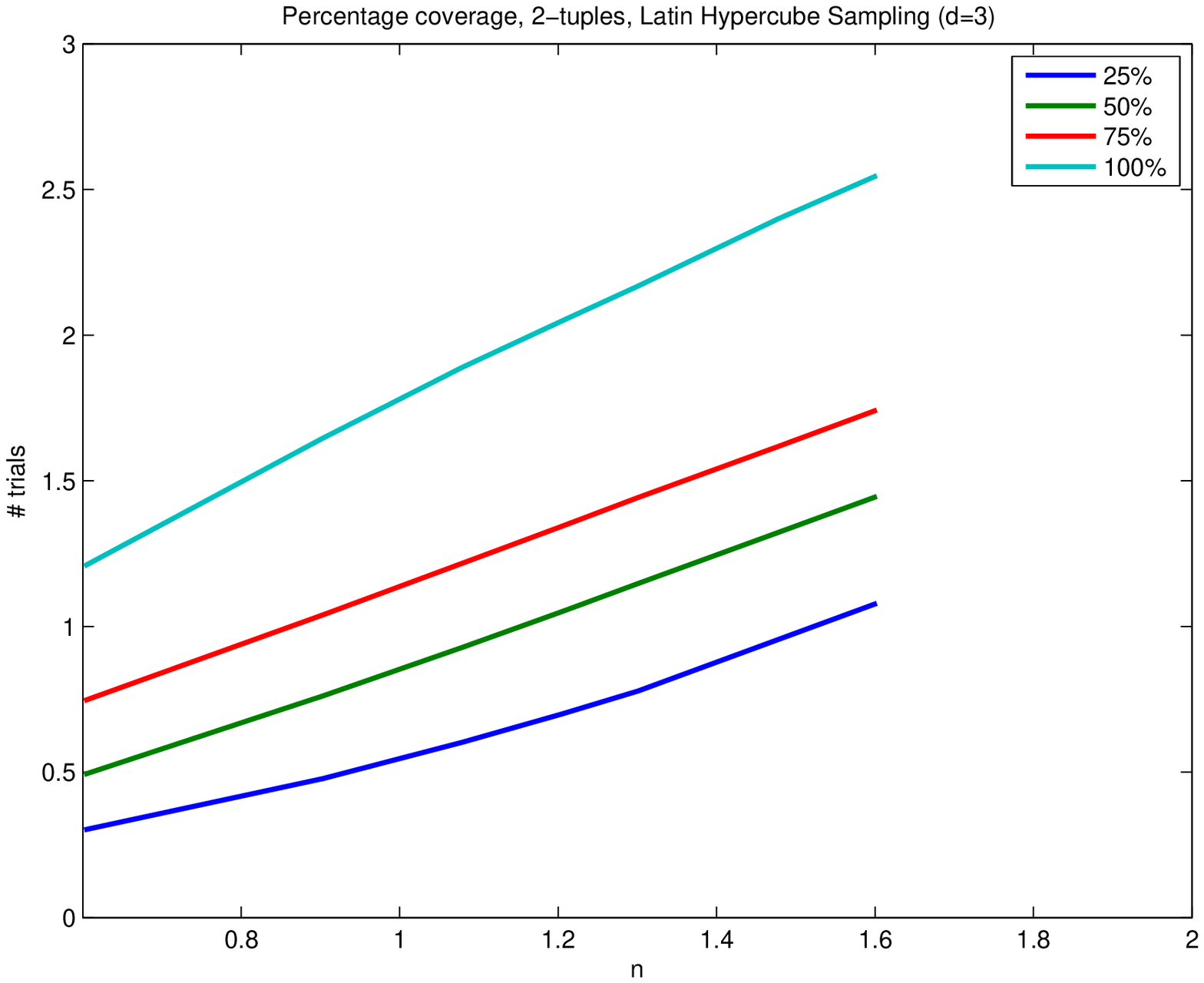} & \includegraphics[width=0.45\textwidth]{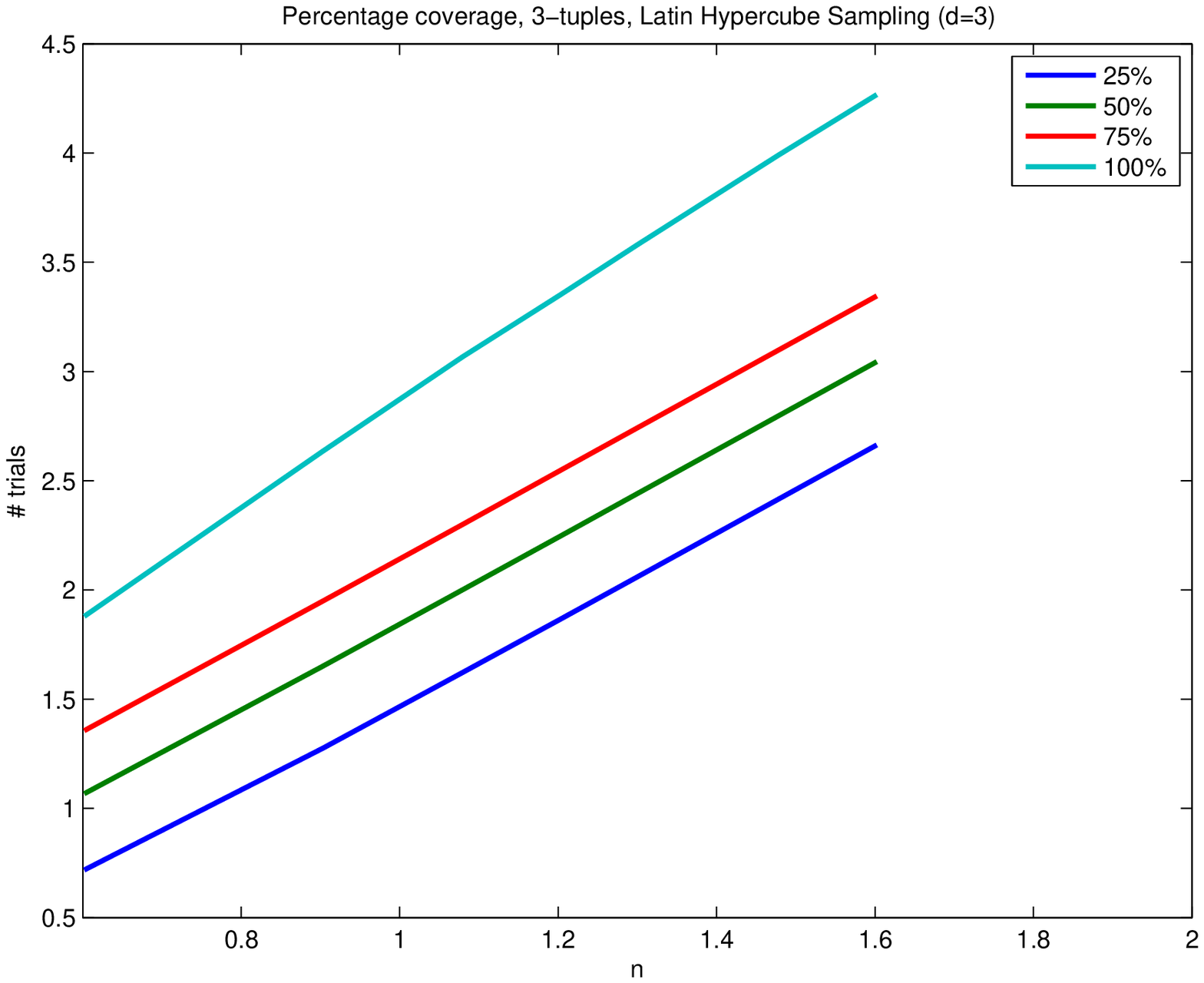}
\label{fig-d3}
\end{array}$$
\end{figure}

\begin{figure}[htb]
\caption{Coverage for 2-tuples (left) and 3-tuples (right), for LHS, $d = 4$}
\centering
$$\begin{array}{cc}
\includegraphics[width=0.45\textwidth]{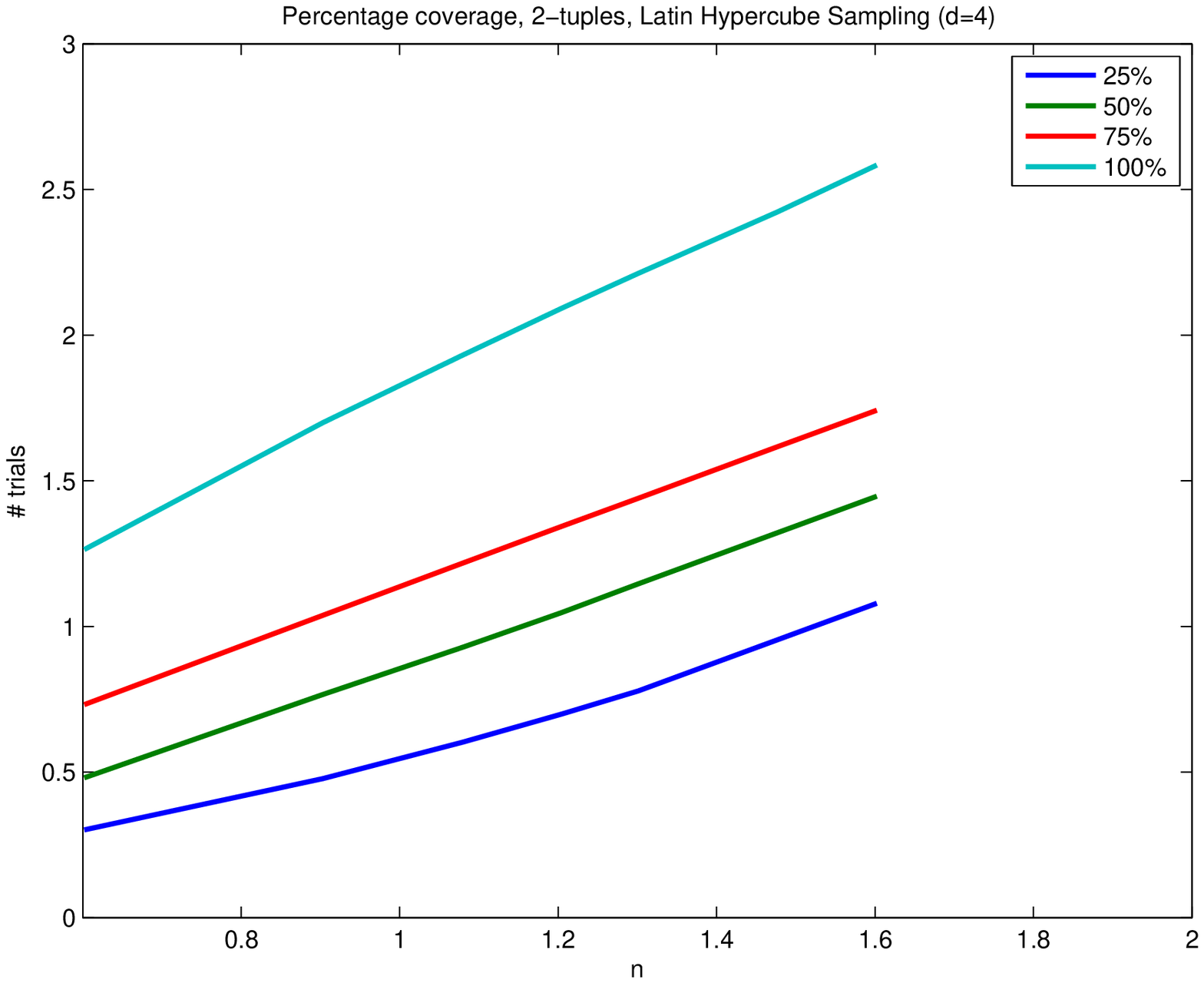} & \includegraphics[width=0.45\textwidth]{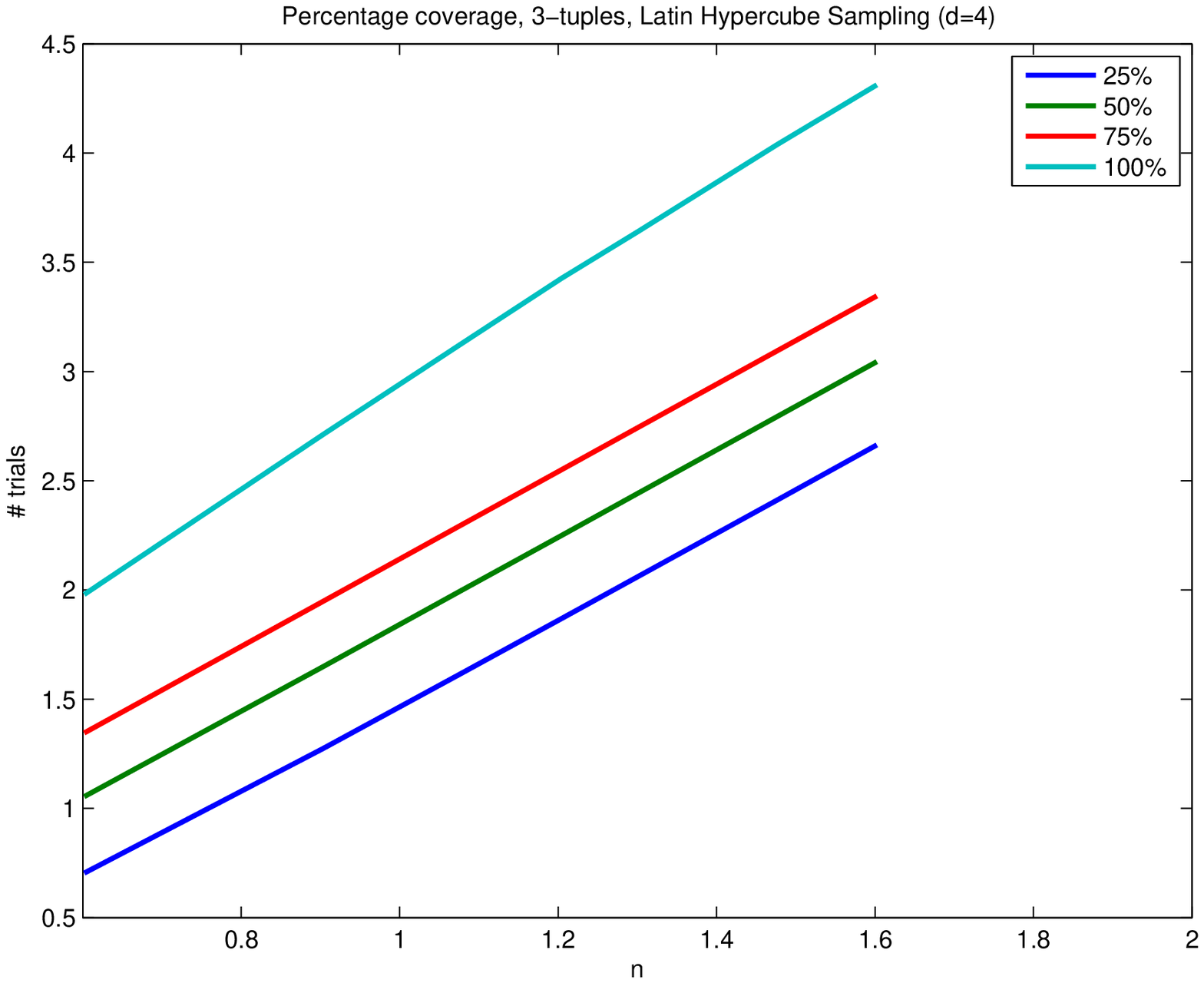}
\label{fig-d4}
\end{array}$$
\end{figure}

\begin{figure}[htb]
\caption{Coverage for 2-tuples (left) and 3-tuples (right), for LHS, $d = 5$}
\centering
$$\begin{array}{cc}
\includegraphics[width=0.45\textwidth]{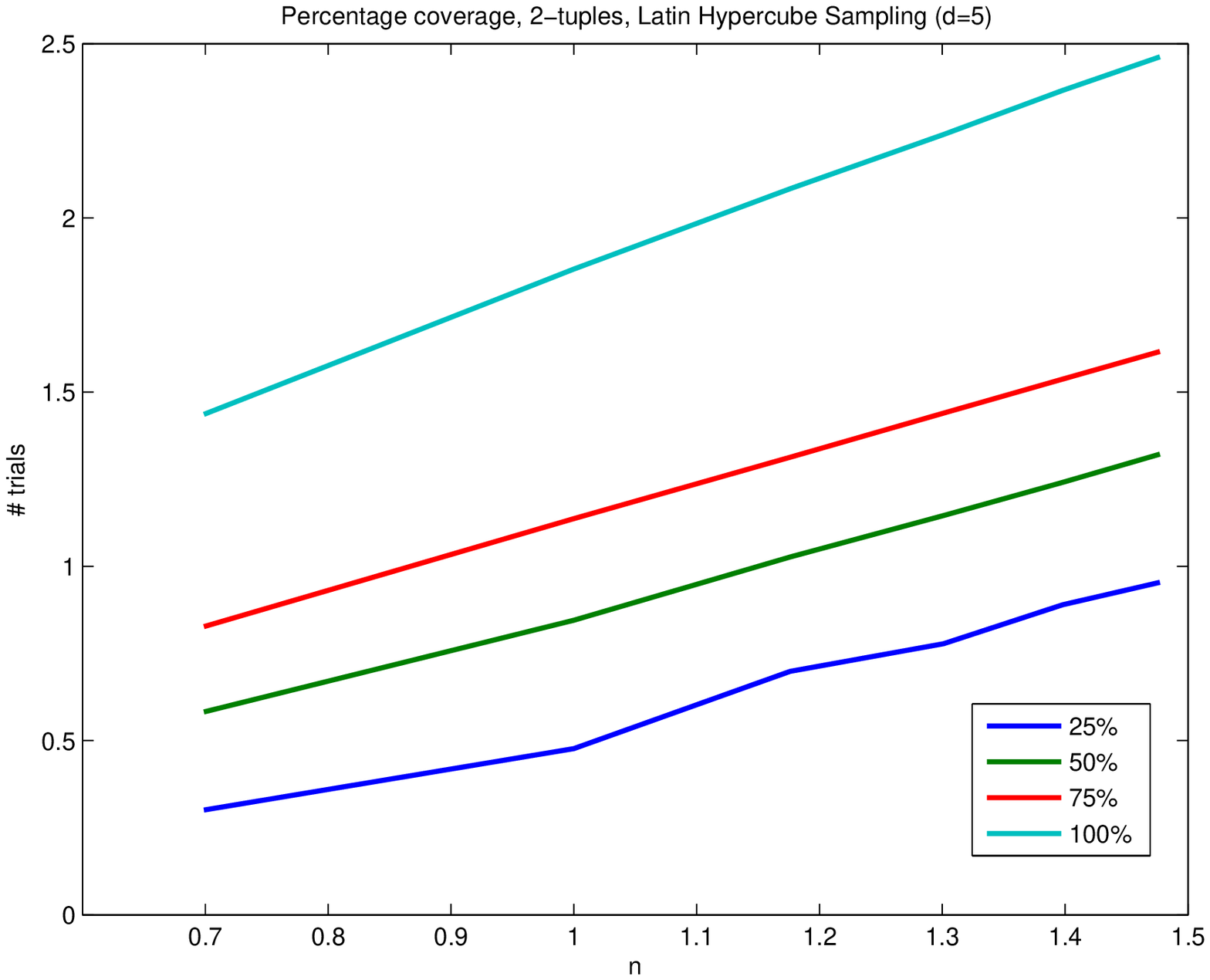} & \includegraphics[width=0.45\textwidth]{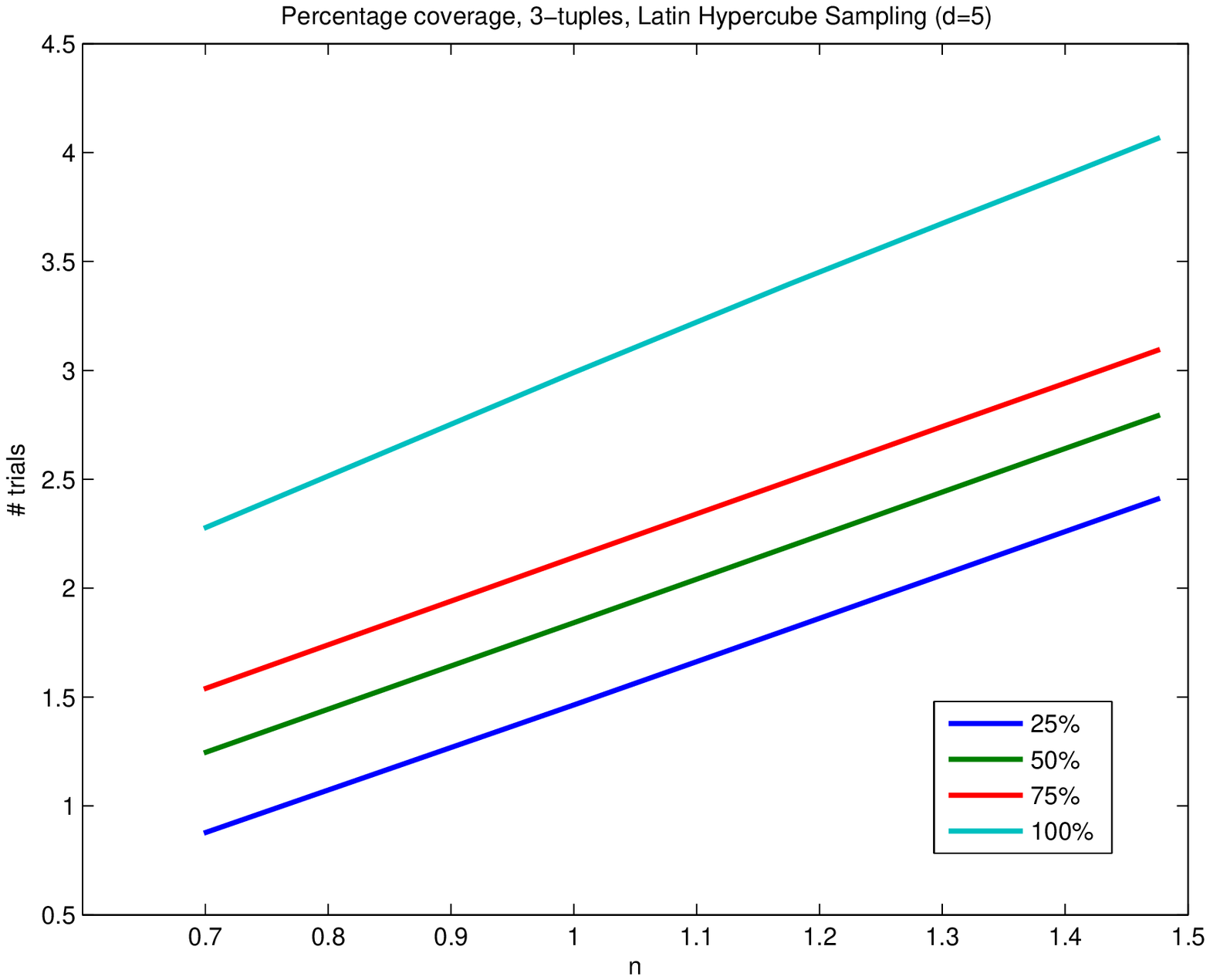}
\label{fig-d5}
\end{array}$$
\end{figure}

\begin{figure}[htb]
\caption{Coverage for 4-tuples for LHS, for $d = 4$ (left) and $d=5$ (right)}
\centering
$$\begin{array}{cc}
\includegraphics[width=0.45\textwidth]{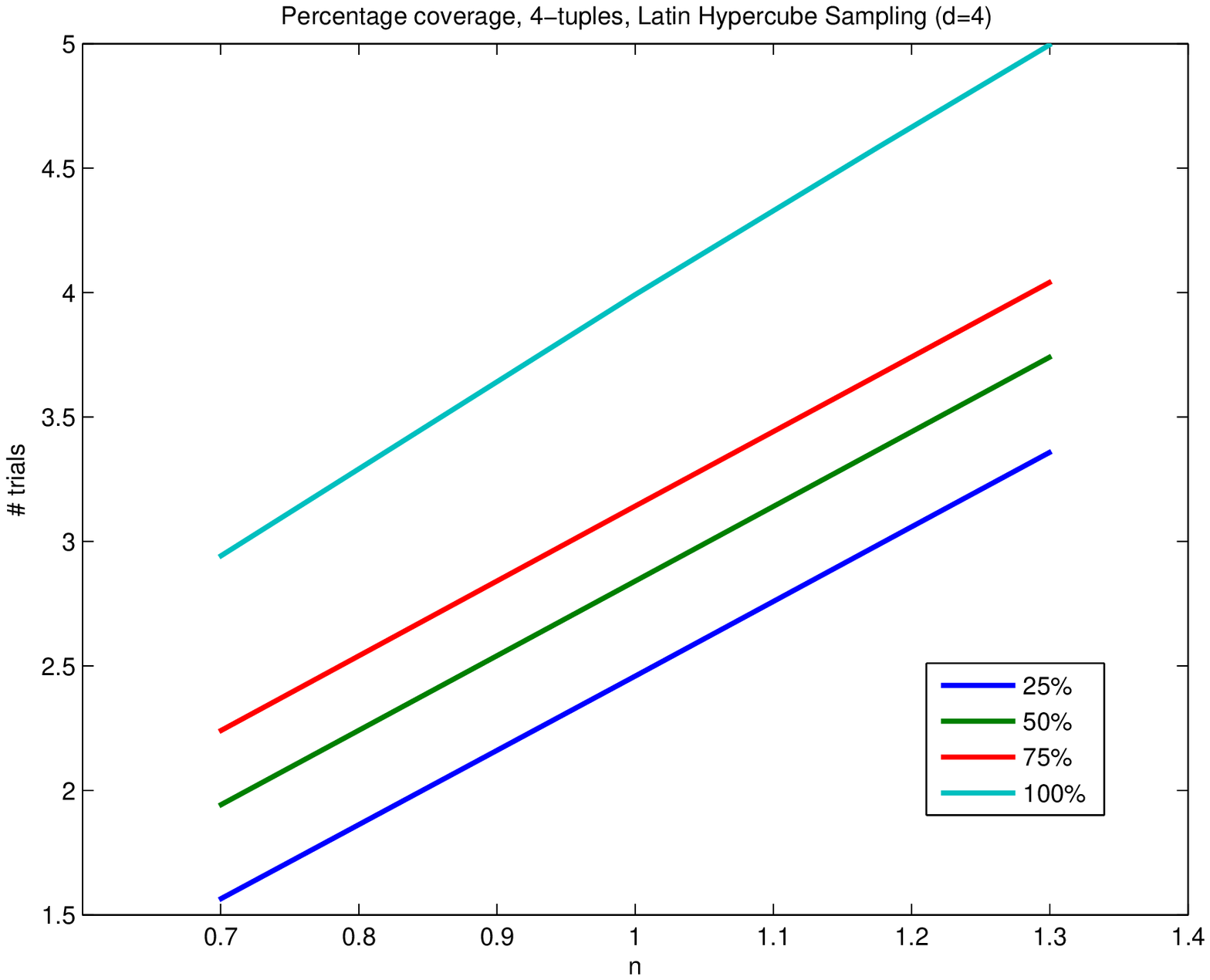} & \includegraphics[width=0.45\textwidth]{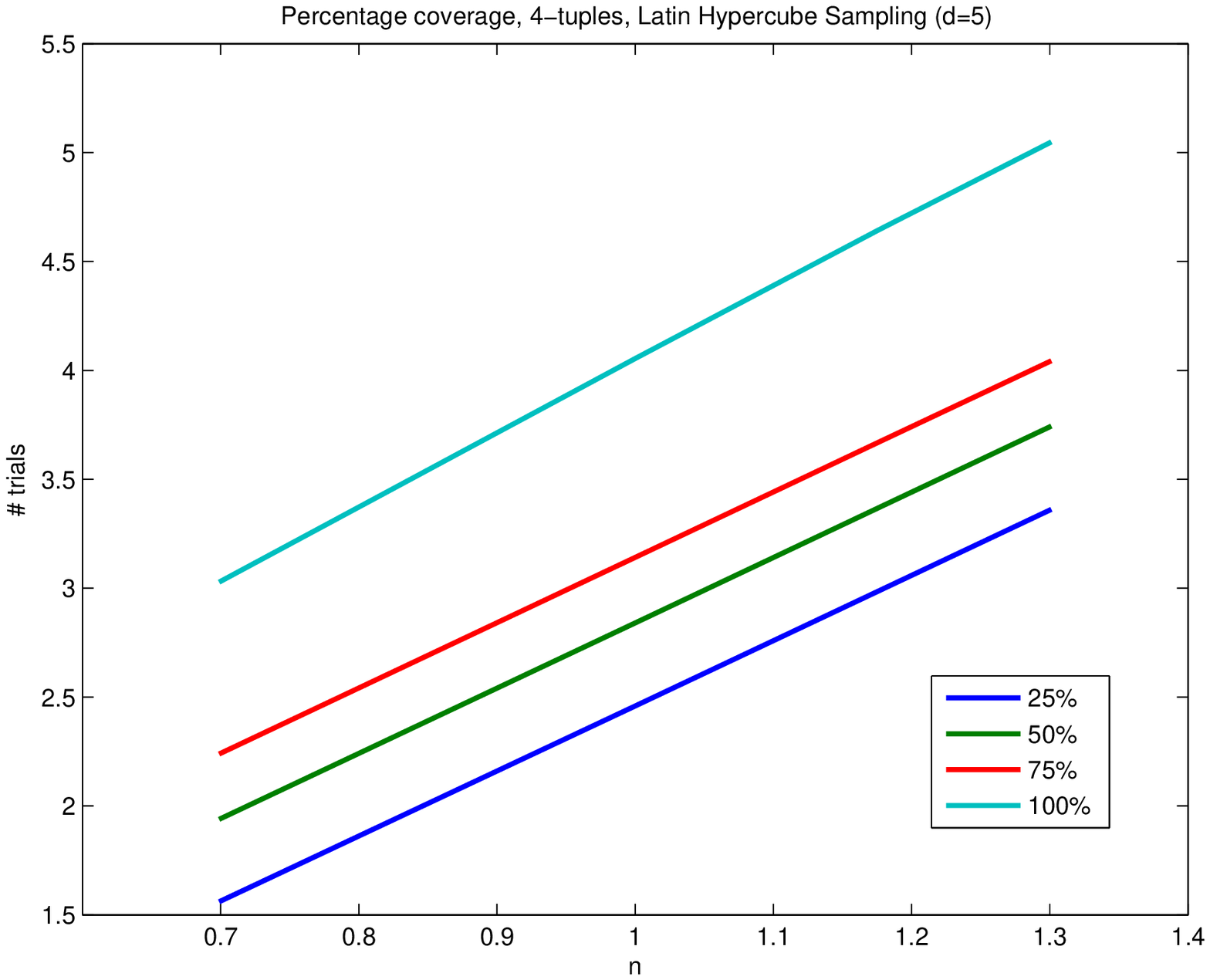}
\label{fig-4tuples}
\end{array}$$
\end{figure}

 In Fig (\ref{fig-d3}) we show LHS results when $d=3$, for 2-tuples and 3-tuples, with coverage 25\%, 50\%, 75\% and 100\%.  Fig (\ref{fig-d4}) shows LHS results when $d = 4$, and Fig (\ref{fig-d5}) gives the $d = 5$ results. All the quantities have been averaged over 200 trials, and the graphs plot the $\log_{10}$ of the data.

When we look at the 2-dimensional subspaces ($t = 2$) with $d=3, 4, 5$, we observe that the number of trials required for a specific percentage coverage is similar, regardless of the dimension $d$ of the system.  In particular, the gradient at 25\%, 50\%, 75\% coverage is 1, while the gradient at 100\% coverage is approximately 1.25 for all values at $d=3, 4, 5$.  We observe a correspondingly similar behaviour for 3D subspaces ($t=3$) for $d=3, 4, 5$, except in these cases the gradient is 2 for incomplete coverage and approximately 2.3 for 100\% coverage.

In \cite{A4} we suggested that in the case $d=2$ the percentage cover for $k$ trials and $n$ divisions is  $1-e^{-k/n}$.  The results here with $t=d=3,$ and $t=d=4$ suggest that the percentage coverage when $t=d$ is given by

$$  P(k,n,t,t)= 1-(1-1/n^{t-1})^k$$
and that in the asymptotic limit as $k$ becomes large then
$$P(k,n,t,t)=1-e^{-k/n^{t-1}}.$$

More generally we conjecture for  any $t<d$ that

$$  P(k,n,d,t)= 1-(1-1/n^{t-1})^k$$
and that in the asymptotic limit as $k$ becomes large then
$$P(k,n,d,t)=1-e^{-k/n^{t-1}}.$$

This is consistent with the 25\%, 50\%, 75\% coverage in which the gradient of the log data is $t-1$.  The only question to address is why the gradient is slightly larger than $t-1$ for 100\% coverage.  To see this we see that 100\% coverage implies $P(k,n,d,t) >1-1/n^{t-1}$.  Thus under our conjecture

$$ 1 - 1/n^{t-1} > 1-(1-1/n^{t-1})^k$$
or

$$ (1-p)^k>p, \quad p=1/n^{t-1}.$$
Using the fact that $\log(1-p) \approx -p $ for $p$ small, then this implies

$$ k \approx (t-1)\log(n) n^{t-1}$$
and so
$$\log(k) \approx (t-1) \log(n)+\log(t-1) +\log(\log(n)).$$
It is this latter term that gives an apparent gradient slightly larger than $t-1$.

Thus we make the following Conjecture

\noindent {\bf Conjecture:} The coverage of  a $t$ dimensional subspace of a $d$ dimensional parameter space of size $n$ when performing $k$ trials of Latin Hypercube sampling is given by $P(k,n,d,t)= 1-(1-1/n^{t-1})^k$ or $1-e^{-k/n^{t-1}}$ when $k$ is large.

Thus if costs and/or experimental factors influence the size of the sample, we can use this information to direct our experiments. So this builds confidence in the modelling results.

\begin{figure}[htb]
\caption{Sub-block coverage in each of the 2-dimensional subspaces for LHS with d=3, n=27, for trials giving 25\% and 75\% coverage.}
\centering
$$\begin{array}{cc}
\begin{array}{c}
\ \\
\includegraphics[width=0.5\textwidth,height=0.4\textheight]{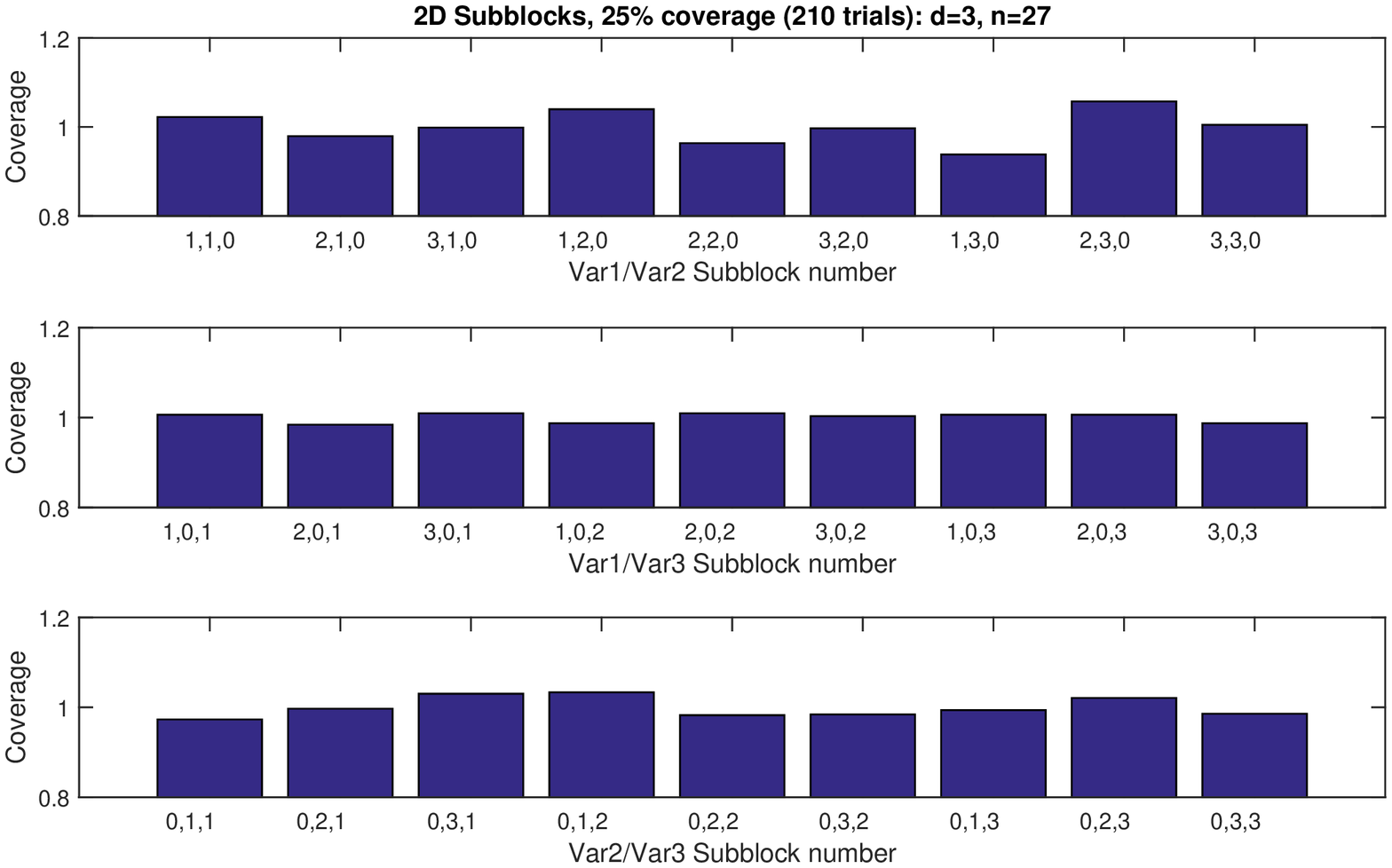}
\end{array} &
\begin{array}{c}
\includegraphics[width=0.5\textwidth,height = 0.4\textheight]{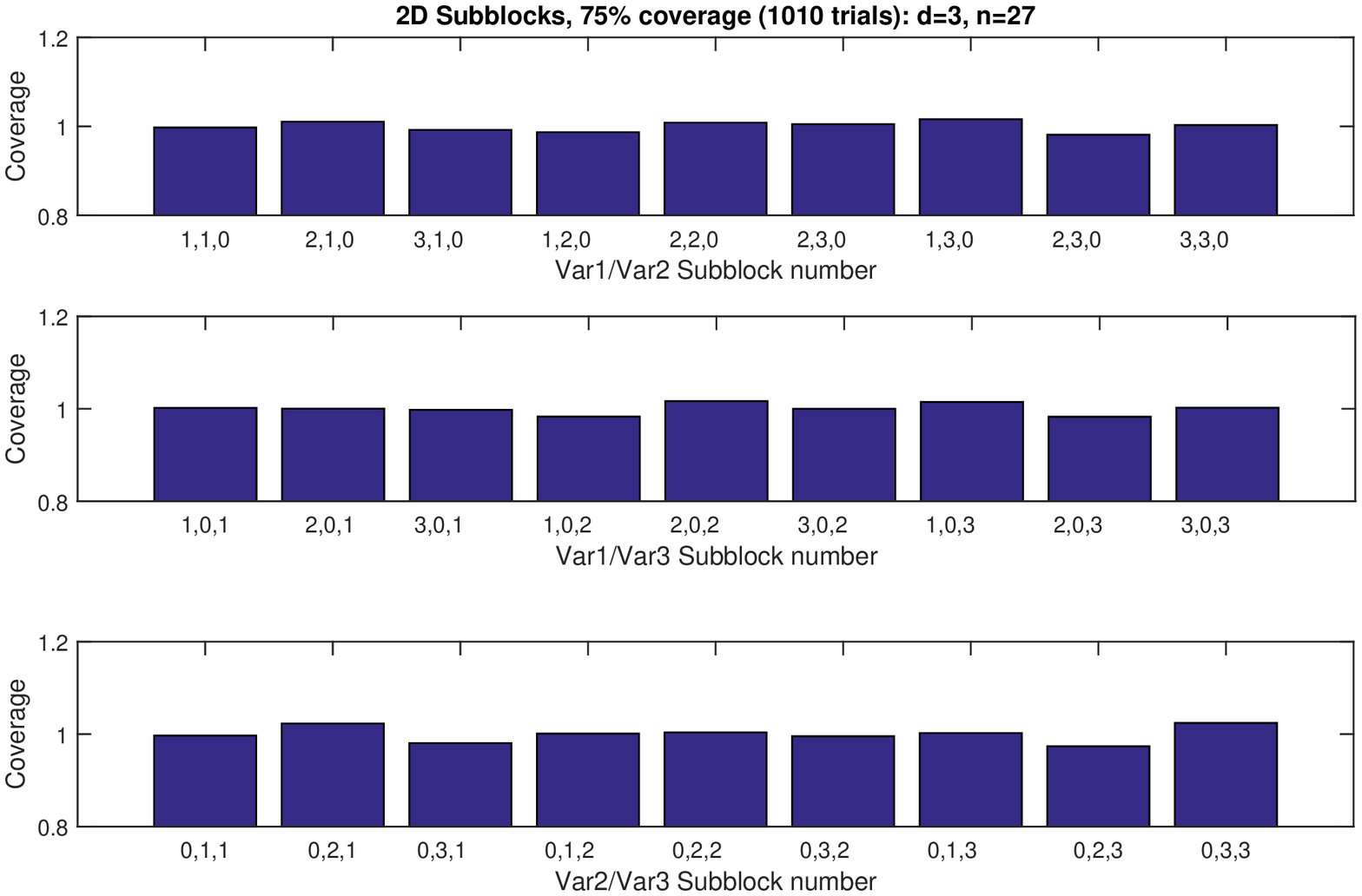}
\end{array}
\label{fig-theory}
\end{array}$$
\end{figure}

 For LHS, where $d=3$ and $n=27$, we investigate the variability (see Fig (\ref{fig-theory})) in coverage of the sub-blocks of the $2$-dimensional spaces, and compare this with Orthogonal sampling where by design the coverage is uniform over the sub-blocks.

The results in the bar graphs can be interpreted by taking a 3-dimensional parameter space, where each of the three variables takes $n=p^d=3^3=27$ distinct levels. Then we partition each 1-dimensional space into $p=3$ sub-blocks of size $p^{d-1}=9$. We are interested in counting the number of points that lie in each $p^{d-1} \times p^{d-1}$ sub-block when projected onto the 2-dimensional subspaces. Our simulations take the average number of Latin Hypercube trials needed to cover 25\% and 75\% of the parameter space and then count the number of points in each of the 2-dimensional sub-blocks. This number is taken as a fraction of the number of trials. For Orthogonal sampling this fraction is 1 across all sub-blocks but as can be seen from the figures, there is much variability when the points are generated using Latin Hypercube trials.

This emphasises  the value of Orthogonal sampling versus Latin Hypercube sampling, where the latter is shown to not cover the sample space uniformly at percentage coverings that are less than 100\%.

 \section{Discussion and conclusions}
 In this paper we have used simulations to give a conjecture about the coverage of  a $t$ dimensional subspace of a $d$ dimensional parameter space of size $n$ when performing $k$ trials of Latin Hypercube sampling.  This coverage takes the form $P(k,n,d,t)=1-(1-1/n^{t-1})^k$ or $1-e^{-k/n^{t-1}}$ when $k$ is large.  This extends the work in \cite{A4}. We suggest that the coverage is independent of $d$ and this allows us to make connections between building Populations of Models and Experimental Designs.  We also show that Orthogonal sampling is superior to Latin Hypercube sampling in terms of giving a more uniform coverage of the $t$ dimensional subspace at  the sub-block size level when only attempting partial coverage of this subspace.  We will attempt to prove our conjecture analytically in a  subsequent paper.




\begin{thebibliography}{50}
\bibitem{A1} O. J. Britton; A. Bueno-Orovio, K. Van Ammel, H.R. Luc, R. Towart, D.J. Gallacher and B. Rodriguez,
Experimentally calibrated population of
models predicts and explains inter subject variability
in cardiac cellular electrophysiology,
 {\em PNAS}, 2014, DOI:10.1073/pnas.1304382110.
 
\bibitem{BC1} R.C. Bryce and C.J. Colbourn,
The density algorithm for pairwise interaction testing,
{\em Software Testing, Verification and Reliability}, {\bf 17}, {\bf 17}, 2007,
  159–-182.


\bibitem{BC2} R.C. Bryce and C.J. Colbourn,
The density-based greedy algorithm for higher strength covering arrays,
{\em Software Testing, Verification and Reliability}, {\bf 19}, 2009, 37–-53.


\bibitem{A4} K. Burrage, P.M. Burrage, D.  Donovan, T.  McCourt and H.B.  Thompson,
Estimates on the coverage of parameter space using populations of models,
 {\em Modelling and Simulation, IASTED, ACTA Press}, 2014,
  DOI: 10.2316/P.2014.813-013.

\bibitem{A3} A. Carusi, K. Burrage and B. Rodriguez,
Bridging Experiments, Models and Simulations: An Integrative Approach to Validation in Computational Cardiac Electrophysiology, {\em Am. J. Physiology}, {\bf 303(2)}, 2012, H144--55.

\bibitem{CD} C. J. Colbourn and J. H. Dinitz, (Eds), Handbook of combinatorial designs, Second Edition,
Chapman \& Hall/CRC,
2006,
  Boca Raton, FL.

\bibitem{A5} C. C. Drovandi, A. N. Pettitt and M. J. Faddy,
Approximate Bayesian computation using indirect inference,
{\em Journal of the Royal Statistical Society: Series C (Applied Statistics)}, {\bf 60(3)}, 2011,
  317–-337.
  
\bibitem{KR} D.R. Kuhn, D.R.Wallace and A.M. Gallo,
Software fault interactions and implications for software testing,
{\em IEEE Transactions on Software Engineering}, {\bf 30(6)}, 2004, 418–-421.
  
\bibitem{MT} E. Marder and A. L.Taylor,
Multiple models to capture the variability of biological neurons and networks,
{\em Computation and Systems, Nature Neuroscience},
   {\bf 14(2)}, 2011, 133--138.
\bibitem{MBC} M.D. McKay, R.J. Beckman and W.J. Conover,
A comparison of three methods for selecting values of input variables in the analysis of output from a computer code,
{\em Technometrics}, {\bf 21(2)}, 1979, 239--245.

\bibitem{Stein} M. Stein, Large sample properties of simulations using Latin hypercube sampling,
 {\em Technometrics}, {\bf 29(2)}, 1987, 143--151.
 
\bibitem{A7} B. Tang, Orthogonal Array-Based Latin Hypercubes,
 {\em Journal of the American Statistical Association}, {\bf 88(424)}, 1993,
1392-–1397.
 
\bibitem{A2} J. Walmsley, J.F. Rodriguez, G.R. Mirams, K. Burrage, I. R. Efimov and  B. Rodriguez,
 MRNA expression
levels in failing human hearts predict cellular
electrophysiological remodelling: A population based
simulation study,
 {\em PLoS ONE}, {\bf 8(2)}, 2013, e56359


\bibitem{A8} W.J. Welch, R. J. Buck, J. Sacks, H.P. Wynn, T.J. Mitchell and M.D. Morris,
Screening, Predicting, and Computer Experiments,
 {\em Technometrics}, {\bf 34}, 1992, 15–-25.
 
\end{thebibliography}

\end{document}